\documentstyle[preprint,aps]{revtex}

\begin{document}

\draft

\title{Null geodesics in the Alcubierre warp drive spacetime: the view 
from the bridge}

\author{Chad Clark \footnote{e-mail: chadc@orion.physics.montana.edu}, 
William A.\ Hiscock\footnote{e-mail: hiscock@montana.edu} and
Shane L.\ Larson\footnote{e-mail: shane@orion.physics.montana.edu}}

\address{Department of Physics, Montana State University, Bozeman, 
Montana 59717}

\date{\today} \preprint{MSUPHY99.02} \maketitle

\begin{abstract}

The null geodesic equations in the Alcubierre warp drive spacetime are
numerically integrated to determine the angular deflection and
redshift of photons which propagate through the distortion of the
``warp drive'' bubble to reach an observer at the origin of the warp
effect.  We find that for a starship with an effective warp speed
exceeding the speed of light, stars in the forward hemisphere will
appear closer to the direction of motion than they would to an
observer at rest.  This aberration is qualitatively similar to that
caused by special relativity.  Behind the starship, a conical region
forms from within which no signal can reach the starship, an effective
``horizon''.  Conversely, there is also an horizon-like structure in a
conical region in front of the starship, into which the starship
cannot send a signal.  These causal structures are somewhat analogous
to the Mach cones associated with supersonic fluid flow.  The
existence of these structures suggests that the divergence of quantum
vacuum energy when the starship effectively exceeds the speed of
light, first discovered in two dimensions, will likely be present in
four dimensions also, and prevent any warp-drive starship from ever
exceeding the effective speed of light.

\end{abstract}

\pacs{ }

Alcubierre \cite{Alcubierre} has described a spacetime which has
features reminiscent of the ``warp drive'' common in science fiction
lore.  The Alcubierre solution allows a ``starship'' to have an
apparent speed relative to distant observers which is much greater
than the speed of light, an effect caused by the spacetime expanding
behind and contracting in front of the starship.  In such a spacetime,
passengers on the starship can travel arbitrarily large distances in
small amounts of proper time; further, there is no time dilation
effect between the starship and clocks outside the region affected by
the warp drive.

If a technology based on such a spacetime could be realized, space
travel to distant points in our Universe could seem almost plausible.
The Alcubierre warp drive spacetime has properties, however, which
make it unlikely to be physical.  As Alcubierre himself pointed out,
in order to create the distortion of spacetime which produces the warp
drive effect, ``exotic'' matter is required, which violates the weak,
strong, and dominant energy conditions.  Although quantized fields can
locally violate the energy conditions, an analysis by Pfenning and
Ford \cite{PF} has shown that the distribution of exotic matter needed
to generate the warp `bubble' around the starship appears quite
implausible.  Additional work has shown that any spacetime that
permits {\it apparent} superluminal travel will inevitably violate the
weak and averaged null energy conditions \cite{Olum,Viss,Low}. On
the other hand, Van Den Broeck has recently shown how to significantly
reduce the amount of negative energy density matter required for the warp
drive to the order of grams \cite{VDB}.

Even if one could somehow obtain the negative energy density matter
needed to support such a spacetime, it appears that quantized fields
may prevent an Alcubierre starship from exceeding the
apparent speed of light.  Hiscock \cite{His2d} has calculated the
vacuum stress-energy tensor of a quantized scalar field in a
two-dimensional reduction of the Alcubierre spacetime and shown that
the stress-energy diverges if the apparent speed of the ship exceeds
the speed of light.  The divergence is associated with the formation
of a horizon in the two-dimensional spacetime. 

Despite the apparent technical difficulties involved in the
construction of an Alcubierre warp drive, it is interesting to ask how
the exterior Universe appears to an observer riding a starship at the
center of the warp drive distortion, on the `bridge' of the starship
\footnote{One might argue that the center of the warp bubble would be
in the engineering area of the starship\cite{MS}; for our purposes,
we shall define the center to be the `bridge' of the starship.}.
Our motivations for the calculation are twofold.  First, it is simply
a fun question to ask, and of interest to compare aberration and
redshift effects to the familiar case of the view seen by a highly
relativistic observer in Minkowski space.  Second, owing to the
existence of the two-dimensional quantum vacuum energy divergence,
it is of interest to determine whether horizon-like causal structures
form in four-dimensions around the starship
whenever the apparent velocity exceeds the speed of light. 

Another possible calculation that could be done is to study how
photons propagate to observers on the far side of a warp bubble from
the source.  This would determine how a passing warp drive starship
would affect the view of the stars seen by a distant astronomer.  We
have not examined this possibility here; the small angular size of the
warp bubble as seen by a distant astronomer would make it unlikely
that such effects would be observed, unless there are enormous numbers
of warp drive starships plying the Galaxy.

The Alcubierre warp drive spacetime \cite{Alcubierre} is described by 
the metric
\begin{equation}
  ds^{2} = -dt^{2} + \left( dx - v_{s} f(r_{s})\ dt \right)^{2} + 
  dy^{2} + dz^{2} \, \,\, ,
\label{Metric}
\end{equation}
where $v_{s} = dx_{s}/dt$ is the apparent velocity of a spacetime 
distortion (the `warp bubble') propagating along a trajectory 
described by $x_{s}(t)$.  For simplicity we will choose the trajectory 
to lie along the $x$ axis, so that $y_s(t) = z_s(t) = 0$.  A radial 
measure of the distance from the center of the warp bubble is given by 
$r$, defined by
\begin{equation}
r = \left[ (x - x_{s})^{2} + y^{2} + z^{2} \right]^{1/2} \, \, .
\label{r}
\end{equation}
The function $f(r)$ can be any function which is normalized to unit 
value at the center of the spacetime distortion, and falls off rapidly 
at some finite radius, asymptotically approaching zero at large $r$.  
Alcubierre gave a particular example of such a function,
\begin{equation}
f(r) = {{tanh[\sigma(r + R)] - tanh[\sigma(r - R)]} \over {2 
tanh[\sigma R]}} \,\,\, .
   \label{TopHat}
\end{equation}
This sort of function, described as a ``top hat'' provides a region 
around $r=0$ where $f$ is roughly constant, leading to small spacetime 
curvatures (and hence tidal forces) there; then a region where the 
function drops steeply from $f \approx 1$ to $f \approx 0$ around $r = 
R$.  The width of the drop-off region is described by the constant 
$\sigma$.  We shall adopt Alcubierre's choice of $f(r)$ as defining 
the warp drive spacetime.

The null geodesic equations are
\begin{equation}
	p^{\alpha}p_{\beta;\alpha} = 0 \ .
	\label{NullGeodesics}
\end{equation}
Defining $dp^{\alpha}/d\lambda = p^{\beta}p^{\alpha}\,_{,\beta}$, 
where $\lambda$ is an affine parameter measured along the null 
geodesic, allows one to write the geodesic equations of Eq.\ 
(\ref{NullGeodesics}) in conventional differential form by expanding 
the covariant derivative, giving
\begin{equation}
   {{dp^{\alpha}}\over {d \lambda}} + \Gamma^{\alpha}\,_{\mu 
   \nu}p^{\mu}p^{\nu} = 0 \ .
   \label{DifferentialEquations}
\end{equation}
The connection coefficients, $\Gamma^{\alpha}\,_{\mu \nu}$, will 
introduce complicated derivatives of the functions $f(r_{s})$ and 
$v_{s}$ into the differential equations, making them analytically 
intractable for general initial conditions.

If the starship travels along the $x$-axis, the system is 
cylindrically symmetric about that axis.  As a result, the behavior of 
null geodesics that reach the ship at $r=0$ may be completely 
understood in terms of the subset that have the $p^{z}$ component of 
the $4$-momentum equal to zero.  Only two of the geodesic equations 
need then be integrated, and the third nonzero component of the 
$4$-momentum may be obtained through the null normalization condition, 
$p^{\alpha}p_{\alpha} = 0$.  In practice, the three equations for 
$p^t, p^x, p^y$ were numerically integrated, and the null 
normalization of the four-momentum was used as a check on the accuracy 
of the integration.

Evaluating the connection and using the coordinate system of 
Eq.(\ref{Metric}), the $(t,x,y)$ components of 
Eq.(\ref{DifferentialEquations}) take the forms:
\begin{equation}
   {{dp^t}\over {d \lambda}} + \Gamma^{t}\,_{tt}(p^{t})^2 + 
   \Gamma^{t}\,_{xx}(p^{x})^2 + 2\Gamma^{t}\,_{tx}p^{t}p^{x} 
   +2\Gamma^{t}\,_{ty}p^{t}p^{y}+2\Gamma^{t}\,_{xy}p^{x}p^{y} = 0 \ ,
\label{dptdl}
\end{equation}
\begin{equation}
   {{dp^x}\over {d \lambda}} + \Gamma^{x}\,_{tt}(p^{t})^2 + 
   \Gamma^{x}\,_{xx}(p^{x})^2 + 2\Gamma^{x}\,_{tx}p^{t}p^{x} 
   +2\Gamma^{x}\,_{ty}p^{t}p^{y}+2\Gamma^{x}\,_{xy}p^{x}p^{y} = 0 \ ,
\label{dpxdl}
\end{equation}
\begin{equation}
   {{dp^y}\over {d \lambda}} + \Gamma^{y}\,_{tt}(p^{t})^2 + 
   2\Gamma^{y}\,_{tx}p^{t}p^{x} = 0 \ .
\label{dpydl}
\end{equation}

In this paper we will only consider the steady-state problem where the 
warp speed of the starship is constant, so that $v_{s} = v = {\rm 
constant}$ and $x_{s} = v t$.

The null geodesic equations to be integrated 
(Eqs.(\ref{dptdl}-\ref{dpydl})) thus form a set of nonlinear coupled 
ordinary differential equations.  It is also necessary to integrate 
the resulting expressions for $p^{\alpha}$ to obtain the trajectory of 
the null geodesic; only with knowledge of the trajectory can the 
metric connections appearing in Eqs.( \ref{dptdl}-\ref{dpydl})
be evaluated properly to complete the integration of the geodesic.  Thus, 
Eqs.(\ref{dptdl}-\ref{dpydl}) are supplemented with the obvious 
relations
\begin{equation}
   {{dt}\over {d \lambda}} -p^{t} = 0 \ ,
\label{dtdl}
\end{equation}
\begin{equation}
   {{dx}\over {d \lambda}} -p^{x} = 0 \ ,
\label{dxdl}
\end{equation}
and
\begin{equation}
   {{dy}\over {d \lambda}} -p^{y} = 0 \ .
\label{dydl}
\end{equation}

It would be inordinately difficult to fire photons from infinity and 
examine which ones actually pass into the (moving) warp bubble and 
reach an observer on the starship at $r=0$.  To avoid this problem, 
the null geodesic equations were numerically integrated for photon 
trajectories originating on the starship's bridge and propagating 
outward through the warp bubble to infinity.  Since the effects of the 
warp bubble are localized, with the spacetime geometry rapidly 
approaching flat space at radii much greater than the bubble radius, 
the numerical integrations were only extended out to $r \sim 100 
R_{bubble}$.  The resulting null geodesics were then time reversed in 
order to give the view as seen from the starship.

The initial conditions at the bridge ($r=0$) consist of choosing the 
photon energy to be of unit value and defining the other components of 
the four-momentum in terms of the bridge angle of the photon via 
Eq.(\ref{bridgeangle}), so that
\begin{eqnarray}
	p^{t} &=& 1 \ , \nonumber \\
	p^{x} &=& \cos(\theta_{0})+v \ , \nonumber \\
	p^{y} &=& \sin(\theta_{0}) \ , \nonumber \\
\label{initcond}
\end{eqnarray}
where, again, the cylindrical symmetry about the {\it x}-axis has been 
used to force $p^{z}$ and $z$ to be zero always.

The two effects of particular interest in this investigation are the 
angular deflection of photons and the shift in their energy as they 
propagate into the warp distortion.  The direction of photon 
propagation at infinity can be obtained from a simple ratio of the 
spatial components of the $4$-momentum,
\begin{equation}
	tan (\theta_{\infty}) = {p^{y} \over p^{x}} \ .
	\label{PhotonAngle}
\end{equation}
To determine the angular deflection of photons, one simply compares 
the value of $\theta$ at infinity to the value of the equivalent angle 
at the center of the warp bubble.  The angle observed at the bridge of 
the starship must be measured with reference to an orthonormal tetrad 
$\{e_{\hat \mu}\}$ moving with the starship.  Such a tetrad may be 
defined by:
\begin{equation}
	(e_{\hat 0})^{\alpha} = u^{\alpha} = (1,v,0,0) \ ,
\label{e0}
\end{equation}
\begin{equation}
	(e_{\hat 1})^{\alpha} = (0,1,0,0) \ ,
\label{e1}
\end{equation}
\begin{equation}
	(e_{\hat 2})^{\alpha} = (0,0,1,0) \ ,
\label{e2}
\end{equation}
\begin{equation}
	(e_{\hat 3})^{\alpha} = (0,0,0,1) \ .
\label{e3}
\end{equation}

The angle observed for an incoming photon is then
\begin{equation}
	\tan(\theta_{0}) = {p^{y} \over {p^{x}-v p^{t}}} \ .
\label{bridgeangle}
\end{equation}
Similarly, the photon energy shift can be obtained by comparing the 
value of the $p^{t}$ component of the photon momentum at the center of 
the warp bubble to the value at infinity,
\begin{equation}
	{E_{0} \over E_{\infty}} = {{p^{t}|_{r = 0}} \over {p^{t}|_{r = 
	\infty}}} \ .
\label{Eratio}
\end{equation}
This measurement is not complicated by the motion of the starship, 
since
\begin{equation}
	E_{0} = -p^{\alpha}u_{\alpha}|_{r = 0} = p^{t}|_{r=0} \ .
\label{Ebridge}
\end{equation}

Analytic results can be obtained for special initial conditions, which 
will provide a useful check of the final numerically obtained general 
results.

Consider a null geodesic originating at $r = 0$ at $90^{\circ}$ to the 
direction of travel.  In this case it is easy to show that an exact 
solution of the null geodesic Eqs.(\ref{dptdl}-\ref{dpydl}) is given 
by
\begin{eqnarray}
	p^{t} &= 1 \ , \nonumber \\
	p^{x} &= v f \ , \nonumber \\
	p^{y} &= 1 \ .  \nonumber \\
\label{orthonullgeo}
\end{eqnarray}
By Eq.(\ref{PhotonAngle}), this photon trajectory is orthogonal to the 
direction of motion at both the bridge and infinity for all warp 
speeds $v$, and by Eq.(\ref{Eratio}), the energy of such a photon is 
the same at infinity and at $r = 0$, and is also independent of the 
warp speed.

Next, consider photons that are aligned with the direction of motion 
of the starship, at either $0^{\circ}$ or $180^{\circ}$.  For such a 
photon, by symmetry, $p^{y}=p^{z}=0$ always.  The null geodesic 
equations for the remaining two components, $(p^{t},p^{x})$ then 
simplify, becoming
\begin{equation}
	{dp^{t} \over d\lambda} + v f_{,x} (p^{t})^2 = 0 \ ,
\label{dptdlzero}
\end{equation}
\begin{equation}
	{dp^{x} \over d\lambda} + v^2 f_{,x} (p^{t})^2 = 0 \ .
\label{dpxdlzero}
\end{equation}
These equations together imply that $p^{x}-vp^{t}$ is a constant of 
the motion.  The existence of this constant, together with the null 
normalization of the photon four-momentum allows the complete solution 
to be written in algebraic form in this case:
\begin{equation}
	p^{t} = {{E_{\infty} (v \pm 1)} \over {v(1-f) \pm 1}} \ ,
\label{ptzero}
\end{equation}
\begin{equation}
	p^{x} = {{E_{\infty} (v \pm 1)(vf \mp 1)} \over {v(1-f) \pm 1}} \ 
	,
\label{pxzero}
\end{equation}
where the upper sign refers to a photon traveling in the opposite 
direction to the ship, so that it meets it head-on, and the lower sign 
refers to a photon traveling in the same direction as the ship, 
chasing it from astern.  Eq.(\ref{ptzero}) allows us to determine the 
energy of these photons as measured at $r = 0$, on the bridge, where $ 
f =1$.  There,
\begin{equation}
	{E_{0} \over E_{\infty}} = 1 \pm v \ .
\label{energyzero}
\end{equation}
We see, as one would expect, that photons running head-on into the 
starship undergo a large blueshift to higher energies, while those 
chasing the starship from astern undergo a redshift.  In fact, as the 
apparent velocity reaches unity, the photons directly astern are 
redshifted to zero energy, and never reach the starship.  This is 
related to the existence of an event horizon in the $(t,x)$ two 
dimensional spacetime when $v \geq 1$, as discussed in detail in 
\cite{His2d}.  

The set of differential equations described by Eqs.\ 
(\ref{dptdl}-\ref{dpydl}) and Eqs.(\ref{dtdl}-\ref{dydl}) were 
numerically integrated using a fourth-order Runge-Kutta routine with 
adaptive stepsize.  The initial conditions were specified at the 
center of the warp bubble, and the null geodesic equations integrated 
outward.  The results of these integrations were then time-reversed 
to determine the appearance of a distant star field to an observer 
riding in the center of the distortion, on the ``bridge'' of the 
``starship''.  The equations were integrated in $2$ degree increments, 
for a variety of constant warp speeds $v$.  The results are most 
easily interpreted in graphical form.  All results illustrated here 
are for $\sigma = 1$; we have examined numerous other cases, and find 
that the qualitative behavior of the null geodesics is largely independent
of the value of $\sigma$.

Figure 1 shows the angular deflection of photons which propagate into 
the warp bubble.  The horizontal axis shows the angle of the photon's 
trajectory to the positive $x$-axis at infinity, while the vertical 
axis shows the angle of the photon's trajectory as measured by an 
observer on the bridge of the starship at $r=0$ (the center of the 
warp bubble).  Note that both angles are defined in terms of the 
apparent direction to the source of the photons (e.g., stars), not in 
terms of the direction of the photon's motion.  For angles at infinity 
of less than 90 degrees, the stars' positions appear to be moved 
towards the direction of motion of the starship---i.e., the angle seen 
at the bridge is less than the angle at infinity.  The magnitude of 
the effect grows with increasing warp speed.  This clustering of 
apparent sources around the direction of motion is qualitatively 
similar to the well-known effect of the aberration of starlight in 
special relativity, illustrated by the two dashed curves (a 
particularly thorough review of the special relativistic visual
effects is provided in Ref.\cite{MD}).

As the angle at infinity increases to 90 degrees and beyond, however, 
new effects which have no analog in special relativity are 
encountered.  First, as shown analytically above, photons which 
originate precisely at 90 degrees to the direction of motion suffer no 
aberration whatsoever, for any warp speed.  A star at 90 degrees to 
the direction of motion will always appear to be at 90 degrees as seen 
from the starship.  Second, there is a conical shaped region behind 
the starship from which no photons reach the starship.  This region 
first forms at warp speed $ v = 1$, at which point a photon from 
directly behind the starship cannot catch up to the starship.  As the 
warp speed is increased, the region behind the starship from which no 
photon can reach the starship grows in angular size, asymptotically 
approaching 90 degrees as the warp speed approaches infinity.  Despite 
the large angular extent of this region, observers on the starship see 
photons originating from all angular directions, as shown in Figure 1.  
Starship observers do not see a large black region, since photons from 
smaller angles (at infinity) appear to arrive from directions further 
behind the ship.

This conical region from which no photon (or other signal) can reach 
the starship when $ v > 1$ forms a sort of ``horizon''.  There exists 
a similar ``horizon'' in front of the starship, a conical region 
inside of which no signal can be received {\it from} the starship.  
These horizon-like structures are the four-dimensional generalizations of the 
behavior previously discovered in examining the two-dimensional warp 
drive spacetime \cite{His2d}, obtained by setting $y = z = 0$.  The 
two-dimensional warp-drive spacetime contains event horizons 
surrounding the starship whenever $v \geq 1$.  Figure 2 illustrates 
how the half-angle of the horizon-like structure grows with increasing warp 
speed. These structures are somewhat analogous to the familiar Mach cones
associated with supersonic fluid flow.

What then would be the view from the bridge on a warp-drive starship
traveling through our galaxy?  The view is best understood by plotting
the energy ratio of the photons versus the angle observed at the
bridge, as is shown in Figure 3, while keeping in mind the aberrations
illustrated in Figure 1.  Assuming the warp drive speed is at least
several times the speed of light, aberration would concentrate an
isotropic distribution of sources along the axis of motion, the
highest density of sources in a small cone directly ahead of the
starship.  Light from these objects would be significantly
blue-shifted, most likely beyond the visible; for the photon striking
the ship head-on, $E_{0}/E_{\infty} = v +1$, as was shown analytically
above.  Again, photons arriving from sources perpendicular to the
direction of motion, at 90 degrees, are unaffected by the warp bubble,
and are observed to have $E_{0}/E_{\infty} = 1$, regardless of the
value of $v$.  Photons coming from the band between 90 degrees and the
edge of the ``horizon'' are spread out by aberration to appear to fill the
hemisphere behind the ship, with a redshift which increases without
bound as the angle at the bridge approaches 180 degrees.  Directly in
front of the starship, blueshifting of incoming photons could create a
hazard for the crew.  Beginning at a warp speed of around $v = 200$,
Cosmic Background photons in the forward hemisphere will be
blue-shifted to energies equivalent to the solar photosphere.

More important than the view is the fact that these calculations have 
shown that the horizon-like structure first discovered in the 
two-dimensional warp drive spacetime exists in the full 
four-dimensional spacetime for any warp speed $v$ that exceeds the 
apparent speed of light.  In the two-dimensional spacetime, it was 
shown that the vacuum stress-energy of a quantized massless field 
would diverge on the horizon as it formed.  Since there exist 
horizon-like structures in the full four-dimensional spacetime, it 
seems likely that a similar divergence will occur over the entire area 
of that structure.  If so, then an infinite amount of energy would be 
involved in the divergence, and semiclassical backreaction effects 
would likely prevent the starship from ever attaining an effective 
warp speed greater than the speed of light.  This quantum 
instability would prevent the operation of a warp drive vehicle even 
if one could overcome the significantly stringent requirements on the 
exotic matter needed to create such a spacetime \cite{PF,VDB}.

It should be pointed out that while it appears plausible that the
two-dimensional divergence would extend over the full horizon-like
structure in four dimensions, this has not yet been demonstrated by an
explicit quantum field theory calculation. Further, there remains a
possibility that the quantum state of the field might be manipulated
in such a way as to avoid the divergence.

This research was supported in part by National Science Foundation 
Grant No.  PHY-97348348.

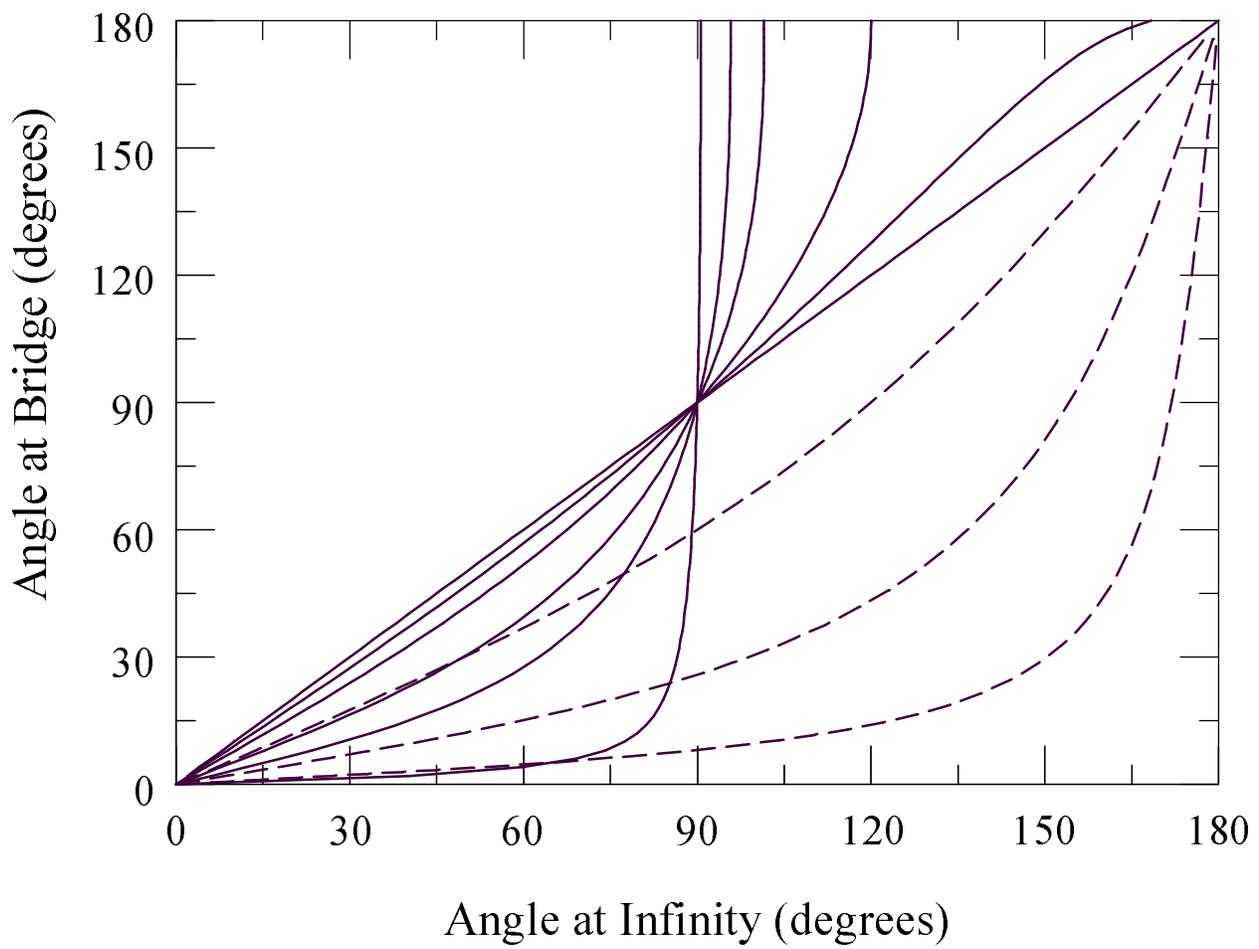
\begin{figure}
\caption{ The apparent angle to a photon source as seen by an observer 
on the starship bridge is compared with the angle as seen by an 
observer at infinity, well outside the warp bubble.  Zero degrees 
corresponds to the direction of motion of the starship.  The solid curves 
represent warp speeds of 0, 1, 2, 5, 10, and 100, from top to bottom 
in the left half of the graph. The dashed curves represent, for comparison,
the special relativistic aberration for a ship traveling at speeds of
0.5, 0.9, and 0.99, from top to bottom}
\end{figure}

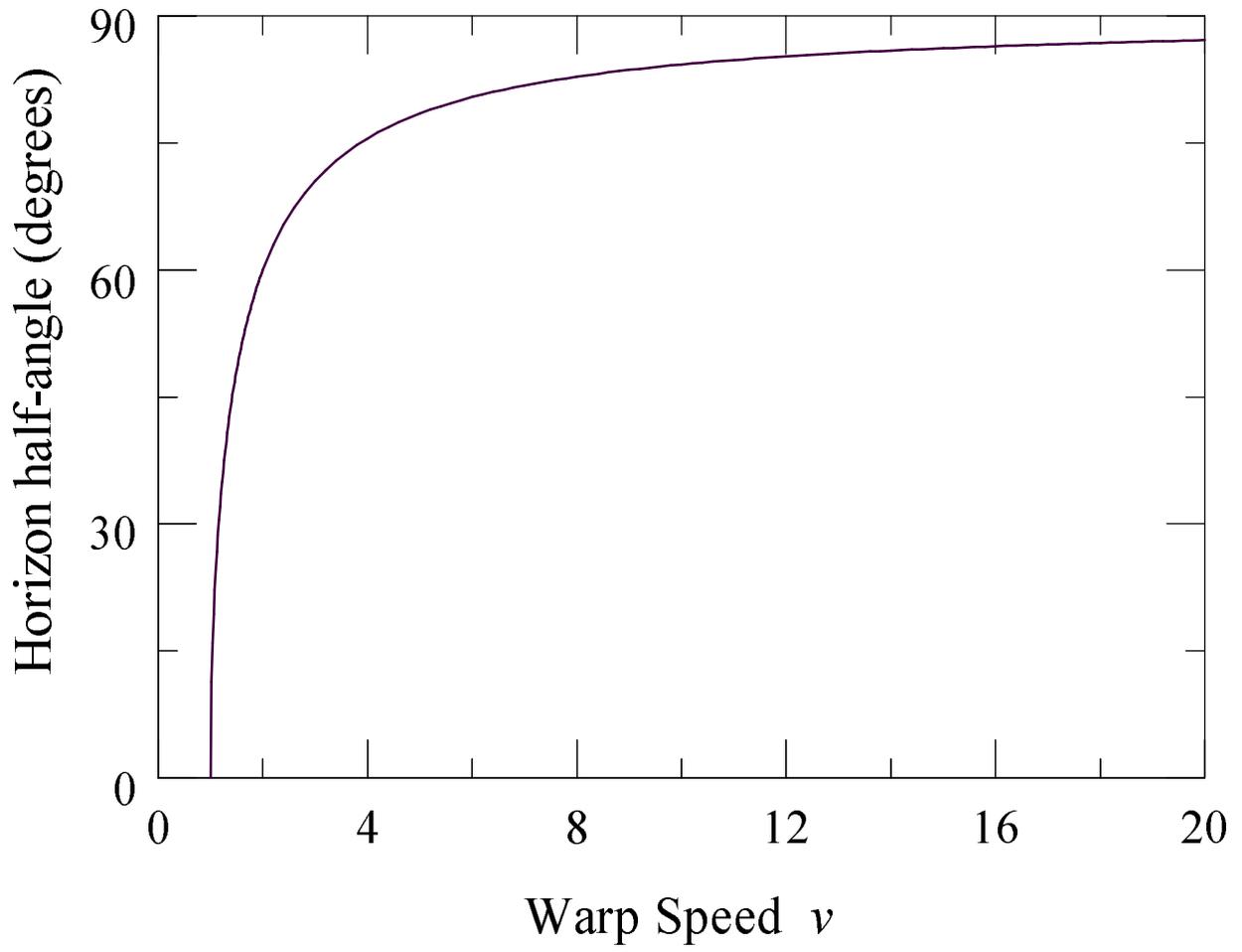
\begin{figure}
\caption{ The half-opening angle of the horizon-like ``cap'' behind the 
starship is shown as a function of the warp speed.  This structure forms 
at warp speed $v =1 $ and grows towards a half-opening angle of 90 
degrees as the warp speed diverges.}
\end{figure}

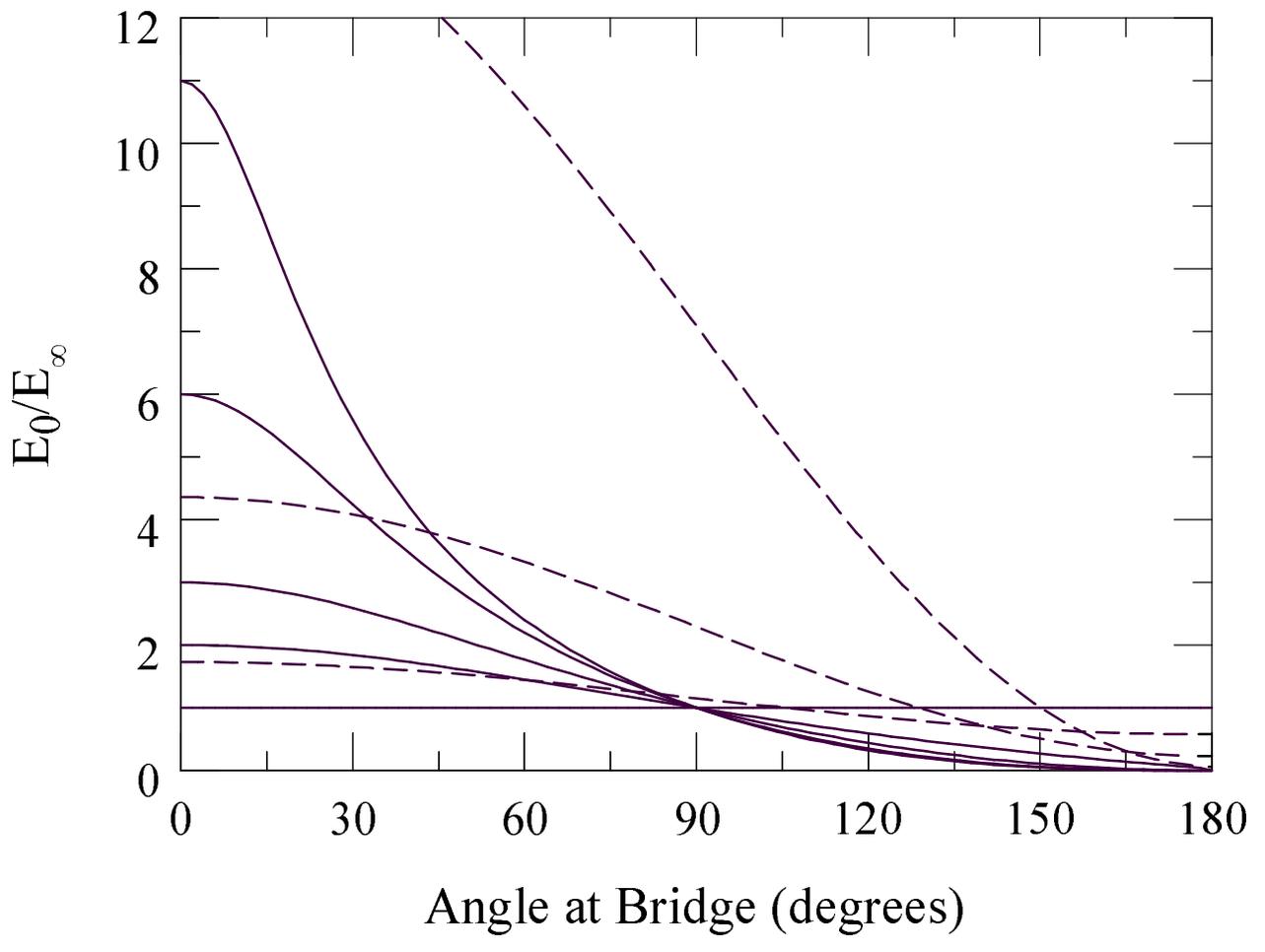
\begin{figure}
\caption{ The ratio of the energy of a photon as measured by an 
observer on the starship bridge to its energy at infinity is shown as 
a function of the angle to the photon source as seen by the observer 
on the starship bridge.  Photons striking the starship head-on are 
blueshifted, their energy increased by a factor $v+1$.  Photons at 90 
degrees are unaffected by the warp bubble.  Photons coming from behind 
the starship are redshifted, with the redshift diverging for photons 
appearing to come from directly behind the ship.  The solid curves represent 
warp speeds of 0, 1, 2, 5, and 10, from bottom to top at the left edge 
of the graph. For comparison, the dashed curves represent the special 
relativistic doppler effect for a ship traveling at speeds of 0.99, 0.9, 
and 0.5, from top to bottom at the left.}
\end{figure}

\end{document}